\documentclass[aps,prl,superscriptaddress,nofootinbib]{revtex4-2}
\usepackage{amsmath,amssymb}
\usepackage{bm}
\usepackage{hyperref}
\usepackage{xcolor}
\usepackage[normalem]{ulem}
\newcommand{\gradf}{\nabla\varphi}
\newcommand{\Av}{\mathbf{A}}
\newcommand{\Bt}{\mathbf{B}}
\newcommand{\Ct}{\mathbf{C}}
\newcommand{\Mv}{\mathbf{M}}
\newcommand{\Nt}{\mathbf{N}}
\newcommand{\Grav}{\mathcal{G}}
\newcommand{\Pg}{\mathbf{P}^{\mathrm{g}}}
\newcommand{\pg}{p_{\mathrm{g}}}
\newcommand{\dd}{\mathrm{d}}
\DeclareMathOperator{\tr}{tr}

\begin{document}

\title{Thermodynamically consistent relaxation dynamics for
	higher-gradient Newtonian gravity}

\author{M. Pszota} \email{Corresponding author, pszota.mate@wigner.hun-ren.hu}
\affiliation{E\"otv\"os Lor\'and University, Budapest, Hungary}
\affiliation{Department of Theoretical Physics, Wigner Research Centre for
	Physics, H-1525 Budapest, Hungary}
\author{P. V\'an}
\affiliation{Department of Theoretical Physics, Wigner Research Centre for
	Physics, H-1525 Budapest, Hungary}
\affiliation{Department of Energy Engineering, Budapest University of
	Technology and Economics, H-1111 Budapest, Hungary}
\affiliation{Montavid Thermodynamic Research Group, Budapest, Hungary}

\date{\today}

\begin{abstract}
	Higher-gradient modifications of Newtonian gravity remove the point-mass singularity. Their static field equations, the regular bi-Yukawa potential and the micrometre-scale laboratory bound are known; a covariant dynamical completion, however, introduces higher time derivatives and the Ostrogradsky instability. We show that the Second Law of thermodynamics, applied to a self-gravitating fluid with a third-order weakly nonlocal state space, provides a different method. Then the potential obeys a first-order relaxation equation, a gradient flow of the gravitational energy whose stationary states are the known sixth-order field equations and to which Ostrogradsky's theorem does not apply. The derivation yields the pressure tensor, whose skew part is an exact divergence, and an off-shell identity that leads to Newton's force law.
\end{abstract}

\maketitle

\emph{Introduction.} Newtonian gravity has no intrinsic length scale and a
divergent point-mass potential. Gradient extensions repair this: adding
first and second gradients of the field strength to the energy yields a
sixth-order field equation whose point-mass solution is finite at the
origin, with two internal lengths~\cite{Lazar2020,LazarWM2019}, in close
analogy with Bopp--Podolsky electrodynamics~\cite{Podolsky1942} and gradient
elasticity \cite{Aif92a,Ber23b}. The price is well known. Made dynamical in the obvious covariant Lagrangian way, $\Delta\to\Box$, such a theory acquires higher time derivatives and falls under Ostrogradsky's theorem~\cite{Woodard2015,PaisUhlenbeck1950}: the Hamiltonian is linear in one of the momenta, the extra pole is a negative-energy mode, and the vacuum is unstable, as in Stelle's higher-derivative gravity~\cite{Stelle1977}. Known escapes operate at the quantum level~\cite{Donoghue2021,LeeWick1969}, through bounded alternative Hamiltonians or conserved energies of the same equations of motion~\cite{Stephen2008,BolonekKosinski2005,Mostafazadeh2010,Kaparulin2014}, or with purely spatial higher derivatives~\cite{Horava2009}.

Here we take a different route. We treat the gravitational potential as a thermodynamic variable of a self-gravitating fluid and derive its evolution from the {Second Law}, extending the construction of~\cite{vanabe,SzucsVan25m} to a third-order weakly nonlocal state space. The result is threefold. First, the higher-gradient field equations of Ref.~\cite{Lazar2020} are recovered as the stationary limit of a \emph{first-order relaxation} equation which is a gradient flow of the gravitational energy: there is no propagating mode, hence no ghost, and stability follows from the concavity of the entropy. Second, the construction determines the pressure tensor and the energy current along with the field equation, and an exact identity returns Newton's force law for the given energy density. Third, the {Second Law} permits a dissipation-free cross-coupling between the field and the bulk viscous pressure of the medium, whose consequences for galactic phenomenology are developed in  ref.~\cite{PszoVan24a}.

The regular point-mass solution, with amplitudes fixed by the ranges, and the micrometre-scale bound on the internal lengths are also due to Ref.~\cite{Lazar2020}; similar cancellations of the Newtonian singularity occur in higher-derivative gravity~\cite{Stelle1978,ModestoNettoShapiro2015,Giacchini2017,AcciolyGiacchiniShapiro2017}. Relaxation and cross-coupling were derived for Newtonian gravity in Refs.~\cite{vanabe,AbeVan22,PszoVan24a,SzucsVan25m}. The novel contributions of this paper are the third-order construction with its pressure tensor, the gradient-flow form with its Lyapunov functional, the stability of the coupled fluid--field system and the divergence form of the skew part of the pressure tensor.

\emph{Balances and state space.} The fluid is described by the mass density $\rho$, velocity $\mathbf v$, specific total energy $e$ and the gravitational potential $\varphi$; an overdot is the substantial derivative. The balances contain no gravitational body force,
\begin{equation}
	\dot\rho+\rho\nabla\cdot\mathbf v=0, ~~
	\rho\dot{\mathbf v}+\nabla\cdot\mathbf P=0,~~
	\rho\dot e+\nabla\cdot \mathbf q=-\mathbf P:\nabla\mathbf v,
	\label{eq:balances}
\end{equation}
gravity entering only through the pressure tensor $\mathbf P$; $\mathbf q$
is the energy flux. The total energy is split as $u=e-\varepsilon/\rho$,
with $u$ the specific internal energy and
\begin{equation}
	\varepsilon=\rho\varphi+\frac{1}{8\pi G}\Bigl[(\gradf)^2
	+\ell_1^2(\nabla^2\varphi)^2+\ell_2^4(\nabla^3\varphi)^2\Bigr],
	\label{eq:eps}
\end{equation}
the gravitational energy density written in the Ohanian form, which is the thermodynamically distinguished representative of the classical energy expressions~\cite{Ohanian2013-ez,trasartibattistoni2026energybalancenewtoniangravitation},
extended by the two higher gradient terms of Ref.~\cite{Lazar2020} with corresponding internal lengths $\ell_1,\ell_2$. The entropy is $s(u,\rho)$, with the usual Gibbs relation defining $T$ and the thermostatic pressure $p$. The Second Law reads $\rho\dot s+\nabla\!\cdot\!\mathbf j\ge0$.

The construction is of weakly nonlocal nonequilibrium thermodynamics:
six fields, six balances, and the constitutive functions
$(\mathbf P,\mathbf q,\mathbf j,\dot\varphi)$ are fixed by the entropy inequality, where the entropy and energy functions $s$, $\varepsilon$ are the only inputs. All model dependence resides in
Eq.~\eqref{eq:eps}.
The applicability of thermodynamics to fields depends on the usual, $\rho$ independent gravitational energy density and using $\rho$ as state variable; extensivity is not presupposed but enters through the coupling $\rho\varphi$ in Eq.~\eqref{eq:eps}~\cite{SzucsVan25m}. The field is thus a thermodynamic variable in its own right, and the field-sector results below -- $\Mv$, $\Nt$, $\Grav$, $\Pg$ and Eq.~\eqref{eq:holo} -- remain valid for $\rho\to0$, i.e.\ outside matter.

\emph{Second Law.} Write $\Av=\partial\varepsilon/\partial(\gradf)$,
$\Bt=\partial\varepsilon/\partial(\nabla^2\varphi)$,
$\Ct=\partial\varepsilon/\partial(\nabla^3\varphi)$ -- of rank one, two
and three, symmetric in all indices -- and
\begin{equation}
	\Mv=\Av-\nabla\!\cdot\!\Bt+\nabla^2\!\!:\Ct,\qquad
	\Nt=\Bt-\nabla\!\cdot\!\Ct .
	\label{eq:MN}
\end{equation}
The gravitational thermodynamic force is
\begin{equation}
    \Grav\equiv\nabla\!\cdot\!\Mv-\partial_\varphi\varepsilon
{=-\delta_\varphi\varepsilon}, 
\end{equation}
where $\delta_\varphi$ denotes the
	variational derivative; for Eq.~\eqref{eq:eps},
$4\pi G\,\Grav=\Delta\varphi-\ell_1^2\Delta^2\varphi
+\ell_2^4\Delta^3\varphi-4\pi G\rho$, so $\Grav=0$ is the sixth-order field
equation of Ref.~\cite{Lazar2020}. Three pressures occur: the thermostatic
$p$; the gravitational pressure tensor
\begin{equation}
	\Pg=-(\varepsilon-\rho\partial_\rho\varepsilon)\mathbf I
	+\Mv\!\otimes\!\gradf+\Nt\!\cdot\!\nabla^2\varphi+\Ct\!:\!\nabla^3\varphi,
	\label{eq:Pg}
\end{equation}
with scalar part $\pg=\tfrac13\tr\Pg$; and the viscous pressure tensor
$\bm\Pi\equiv\mathbf P-p\mathbf I-\Pg$, with bulk viscous pressure
$\Pi=\tfrac13\tr\bm\Pi$ and $\bm \Pi^{\rm dev} \equiv \bm \Pi - \Pi \, \mathbf{I}$, both vanishing in mechanical equilibrium. Substituting
Eq.~\eqref{eq:balances} into the entropy inequality and separating
divergences yields the entropy flux $\mathbf j=\mathbf J_{\rm th}/T$, with
$\mathbf J_{\rm th}$ the energy flux corrected by weakly nonlocal terms, and
the production
\begin{equation}
	T\sigma_s=\mathbf J_{\rm th}\!\cdot\!T\nabla\tfrac1T
	+\Grav\,\dot\varphi-\Pi\,\nabla\!\cdot\!\mathbf v
	-\bm\Pi^{\rm dev}\!:\!(\nabla\mathbf v)^{\rm dev}\ge0 .
	\label{eq:prod}
\end{equation}
One can verify Eqs.~\eqref{eq:Pg}--\eqref{eq:prod} either via Liu's procedure or with the separation of divergences.

By Curie's principle the scalar parts may couple. The linear Onsagerian
equations with a conductivity matrix $L$ and their entropy bound are
\begin{equation}
	\begin{pmatrix}\dot\varphi\\-\Pi\end{pmatrix}
	=\begin{pmatrix}l_1&l_{12}\\ l_{21}&l_2\end{pmatrix}
	\begin{pmatrix}\Grav\\ \nabla\!\cdot\!\mathbf v\end{pmatrix},
	\quad
	l_1l_2\ge\tfrac14(l_{12}+l_{21})^2,
	\label{eq:onsager}
\end{equation}
with $l_1,l_2\ge0$; $l_2$ is the bulk viscosity. For a scalar field no
microscopic time-reversal argument enforces $l_{12}=l_{21}$; the antisymmetric part is dissipation-free. The symmetric part contributes to entropy production (\textit{i.e.} it is dissipative), the antisymmetric part can change the dynamics while contributing no entropy production (\textit{i.e.} it is nondissipative). In particular $l_1=0$ --
\emph{gravity itself produces no entropy} -- forces $l_{21}=-l_{12}$, a
purely reactive coupling. Eliminating $\nabla\!\cdot\!\mathbf v$ (possible
for $l_2\neq0$) gives the central result,
\begin{equation}
	\boxed{\;
		\tau\,\dot\varphi=4\pi G\,\ell^2\bigl(\Grav-6K\,\Pi\bigr),\;}
	\label{eq:fieldeq}
\end{equation}
with $\ell^2/\tau :=\det L/4\pi Gl_2$ and $K:=l_{12}/6\det L$ being constructed from physical constants and the Onsager coefficients; only the ratio $\ell^2/\tau$ is fixed, and $\ell$ is unrelated to $\ell_1,\ell_2$. The substantial derivative reduces to $\partial_t\varphi$ in a medium at rest or when the relaxational timescale is much smaller than the characteristic time associated with the material velocity. The potential relaxes towards {the stationary state $\delta_\varphi\varepsilon=0$}, driven additionally by the bulk viscous pressure. The cross-term is fixed by the independent coefficient $l_{12}$, not by $\varepsilon$; in mechanical equilibrium ($\Pi=0$) it is absent, and for the remainder of this Letter we set $l_{12}=l_{21}=0$, so that $K=0$ and
$\ell^2/\tau=l_1/4\pi G$.

\emph{Potential representability and Newton's force.} Taking the
divergence of Eq.~\eqref{eq:Pg}, the $\Mv$ and $\Nt$ terms cancel
identically and one obtains the exact, off-shell identity
\begin{equation}
	\nabla\!\cdot\!\Pg=\rho\,\nabla(\partial_\rho\varepsilon)
	+\Grav\,\gradf .
	\label{eq:holo}
\end{equation}
On shell ($\Grav=0$) the Newtonian force density follows from the pressure tensor alone given the energy density of Eq.~\eqref{eq:eps}, for which $\partial_\rho\varepsilon=\varphi$:
\begin{equation}
	\nabla\!\cdot\!\Pg=\rho\,\nabla\varphi,
	\label{eq:newtonian-force-density}
\end{equation}
with no separate body-force postulate ~\cite{vanabe,Van23a,SzucsVan25m}.
Since also $\partial_\varphi\varepsilon=\rho$, Eq.~\eqref{eq:holo} is
equivalent to the exact relation $\nabla\!\cdot\!\Pg=(\nabla\!\cdot\!\Mv)\,\gradf$.

The tensor $\Pg$ in eq. \eqref{eq:Pg} is not symmetric in
general. Symmetry does not follow from the entropy inequality; it is an additional
requirement for non-polar media. Here the skew part is an exact divergence,
\begin{equation}
	P^{\rm g}_{ik}-P^{\rm g}_{ki}=\partial_jS_{jik}, \qquad 
	S_{jik}=N_{jk}\partial_i\varphi-N_{ji}\partial_k\varphi 
	+2\,(C_{jkl}\partial_i-C_{jil}\partial_k)\partial_l\varphi .
	\label{eq:skew}
\end{equation}
The field thus contributes a
couple-stress flux, the total angular momentum is conserved. A skew-symmetric pressure tensor appeared in case of gravitational theories extended with material coupling ~\cite{SzucsVan25m}.

The reduction of the momentum balance to the mass-point form $\dot{\mathbf v}=-\nabla\Phi$ (e.g.\ $\Phi=\varphi+h$ for homentropic flow) requires more than \eqref{eq:holo}, and the
additional requirements are thermal rather than gravitational: internal
perfectness ($\Grav=0$), mechanical perfectness ($\Pi=0$,
$\boldsymbol\Pi^{\rm dev}=\mathbf 0$), and integrability of the entropic
force. The last is nontrivial because $\nabla p/\rho$ is not a gradient: the
Gibbs relation for the specific enthalpy gives
$\nabla p/\rho=\nabla h-T\nabla s$, leaving the continuum entropic force
$T\nabla s$, which is a gradient precisely when
$\nabla T\times\nabla s=0$~\cite{SzucsVan25m}. The weakly nonlocal sector,
gradient terms included, is itself of gradient form and contributes nothing
here, so the only obstruction is the classical thermal one already present
for an Euler fluid -- equivalently, the absence of baroclinic vorticity
generation.

\emph{Relaxation dynamics and stability.} Two separate statements are needed.
For \emph{statics}, Ostrogradsky's theorem does not apply at all: it
concerns higher time derivatives, while the theory of
Ref.~\cite{Lazar2020} is a purely spatial elliptic problem. The relevant
requirement is thermodynamic: concavity of the entropy in the extended state
space, i.e.\ convexity of $\varepsilon$, which for Eq.~\eqref{eq:eps} reads
$\ell_1^2>0$, $\ell_2^4>0$.

For \emph{dynamics}, the danger is real on the covariant Lagrangian route --
$\Box(1-\ell^2\Box)\varphi=4\pi G\rho$ propagates a negative-energy pole --
but Eq.~\eqref{eq:fieldeq} with $K=0$, in a medium at rest, is a \emph{gradient flow},
\begin{equation}
	\tau\,\partial_t\varphi=-4\pi G\ell^2\,\delta_\varphi\varepsilon ,
	\label{eq:gradflow}
\end{equation}
Its Lyapunov functional is the gravitational energy
$E[\varphi]=\int\varepsilon\,\dd V$ at fixed $\rho$,
\begin{equation}
	\frac{\dd E}{\dd t}=-\frac{\tau}{4\pi G\ell^2}\int(\partial_t\varphi)^2\,\dd V\le0 ,
	\label{eq:lyap}
\end{equation}
provided $\mathbf J_\varphi=\Mv\,\partial_t\varphi+\Nt\!\cdot\!\nabla\partial_t\varphi
+\Ct\!:\!\nabla^2\partial_t\varphi$, the weakly nonlocal part of $\mathbf J_{\rm th}$
in a medium at rest, has no normal component on the boundary; $E$ is convex and
bounded from below for $\ell_1^2,\ell_2^4\ge0$, and its minimizer solves $\Grav=0$.
Linearizing about any background at fixed $\rho$, a perturbation
$\propto e^{i\mathbf k\cdot\mathbf x+\lambda t}$ obeys
\begin{equation}
	\lambda(k)=-\frac{\ell^2}{\tau}
	\bigl(k^2+\ell_1^2k^4+\ell_2^4k^6\bigr)<0
	\quad\text{for all }k\neq0 ,
	\label{eq:spectrum}
\end{equation}
whenever the entropy is concave: the gradient terms \emph{increase}
the damping at large $k$. There is no negative-energy propagating mode
because there is no propagating mode at all -- the potential carries no
independent dynamical degree of freedom, only a relaxation transient, and
the elliptic (Lagrangian) theory is recovered as the constraint limit
$\tau\to0$. The theorem is evaded not by curing its conclusion but by
failing its hypotheses: it presumes a nondegenerate Lagrangian containing
higher \emph{time} derivatives, hence a canonical Hamiltonian linear in one
momentum, whereas \eqref{eq:fieldeq} is first order in time and carries no
Hamiltonian structure at all. What replaces boundedness of the Hamiltonian is \emph{concavity of the entropy}, which also ensures \eqref{eq:spectrum}. This is different in kind from the quantum routes of Refs.~\cite{Donoghue2021,LeeWick1969}, which retain the Lagrangian and repair the instability after quantization.

For the coupled system -- continuity, momentum and
Eq.~\eqref{eq:fieldeq} with $l_{12}=l_{21}=0$ -- gravity enters only through
$\nabla\!\cdot\!\Pg=(\nabla\!\cdot\!\Mv)\,\gradf$. About a homogeneous state at rest
taken on shell ($\Grav_0=0$, the Jeans swindle~\cite{Kiessling2003}) its linear part
is $\rho_0\nabla\delta\varphi$. Barotropic perturbations
$\propto e^{i\mathbf k\cdot\mathbf x+\lambda t}$, with sound speed $c_s$,
longitudinal kinematic viscosity $\nu$, the field relaxation rate
$\gamma=(\ell^2/\tau)k^2/g$ of Eq.~\eqref{eq:spectrum},
$g=(1+\ell_1^2k^2+\ell_2^4k^4)^{-1}$ and $\omega_J^2=4\pi G\rho_0$, obey
\begin{equation}
	\lambda^3+(\nu k^2+\gamma)\lambda^2+(c_s^2k^2+\nu k^2\gamma)\lambda+\gamma\,(c_s^2k^2-\omega_J^2g)=0 .
	\label{eq:cubic}
\end{equation}
Since $a_2a_1-a_0=\nu k^2(c_s^2k^2+\nu k^2\gamma+\gamma^2)+\gamma\omega_J^2g>0$, the
Routh--Hurwitz conditions reduce to $a_0>0$, and the only instability is the
Jeans one,
\begin{equation}
	c_s^2k^2\,(1+\ell_1^2k^2+\ell_2^4k^4)<4\pi G\rho_0 ,
	\label{eq:jeans}
\end{equation}
with the threshold of the static theory, independent of $\tau$, $\ell$ and $\nu$;
relaxation only lowers the growth rate, to $(\gamma\omega_J^2)^{1/3}$ for
$\gamma\ll\omega_J$.

Two instabilities must not be conflated here. The Jeans instability of
	a self-gravitating fluid is \emph{tachyonic}: the wrong sign belongs to the
	mass term, $\omega^2=c_s^2k^2-4\pi G\rho_0$, exactly as in spontaneous
	symmetry breaking. It is physical, self-resolving, and must be retained by
	any theory of gravity. The Ostrogradsky instability is a \emph{ghost}: the
	wrong sign belongs to the kinetic term, the energy is unbounded below, and
	there is no state to settle into. The two occupy opposite ends of the
	spectrum, $k<k_J$ against $k\sim\ell_1^{-1}$, and do not couple at linear 	order. The relaxational completion thus retains the physical
	instability and adds no other.

\emph{Regularity, the amplitude sum rule, and the laboratory.} For the
stationary linear theory,
$\Delta(1-\ell_1^2\Delta+\ell_2^4\Delta^2)\varphi=4\pi G\rho$, partial fractions give the point-mass potential of Ref.~\cite{Lazar2020}
\begin{equation}
	\varphi(r)=-\frac{GM}{r}\Bigl(1+\sum_i\alpha_i e^{-r/\lambda_i}\Bigr),
	\quad \sum_i\alpha_i=-1 ,
	\label{eq:sumrule}
\end{equation}
with $i\in\{1,2\}$, the sum rule being equivalent to regularity at the origin. The amplitudes are therefore \emph{fixed by the ranges}~\cite{Lazar2020}. The two roots of the factorisation
$1+\ell_1^2k^2+\ell_2^4k^4=(1+ak^2)(1+bk^2)$ obey
\begin{equation}
	a+b=\ell_1^2,\qquad ab=\ell_2^4,
	\qquad \lambda_a=\sqrt a,\ \ \lambda_b=\sqrt b ,
	\label{eq:roots}
\end{equation}
so $a,b$ have the dimension of length \emph{squared} and the Yukawa ranges
are their square roots. Taking $a>b$, partial fractions give
\begin{equation}
	\alpha_a=-\frac{a}{a-b}=-\frac{\lambda_a^2}{\lambda_a^2-\lambda_b^2}\le-1,
	\quad
	\alpha_b=-\alpha_a-1\ge0 .
	\label{eq:amps}
\end{equation}
For a single length $\ell_2=0$, hence $b=0$, $\lambda_a=\ell_1$ and
$\alpha=-1$ exactly. This is the decisive structural difference from a generic fifth force, whose amplitudes and ranges are independent, and it makes the theory unusually easy to test. Torsion-balance experiments exclude a single Yukawa term with $|\alpha|=1$ for $\lambda>38.6\,\mu$m at 95\% confidence~\cite{Lee2020} (earlier $56\,\mu$m~\cite{Kapner2007}); for a single length $\alpha=-1$, $\lambda=\ell_1$) this gives~\cite{Lazar2020}
\begin{equation}
	\ell_1\lesssim4\times10^{-5}\ \mathrm{m} .
	\label{eq:bound}
\end{equation}
For two real lengths the published limits still apply. The torque is
linear in the potential, $T=\alpha_aT_1(\lambda_a)+\alpha_bT_1(\lambda_b)$ with
$T_1$ the response to a unit Yukawa term, and Eq.~\eqref{eq:amps} gives
$|T|-T_1(\lambda_a)=b\,[T_1(\lambda_a)-T_1(\lambda_b)]/(a-b)\ge0$ whenever $T_1$
grows with the range, as for plane-parallel periodic test masses. Hence
$\lambda_a\lesssim4\times10^{-5}$~m and $\ell_1\le\sqrt2\,\lambda_a$. The
oscillatory regime below requires a dedicated analysis of the
data~\cite{Perivolaropoulos2017}.
Because the corrections are exponentially confined below $\ell_1$, all
larger-scale gravity is untouched, while \emph{inside} matter and at
interfaces the higher-order equation requires additional boundary
conditions -- the regime where the theory genuinely differs from Newton.

The factorization $1+\ell_1^2k^2+\ell_2^4k^4=(1+ak^2)(1+bk^2)$ has real
roots for $\ell_1^4>4\ell_2^4$ and complex-conjugate roots otherwise, where the correction becomes a damped oscillation~\cite{Lazar2020}. If a static potential arises
from single-particle exchange with a non-negative spectral density, it is completely monotone~\cite{Widder1941,FeinbergSucher1968} and cannot change sign; the oscillatory regime therefore requires complex-conjugate poles, the classic Lee--Wick situation~\cite{LeeWick1969}. In the present framework this is no obstruction: both stability conditions -- convexity and Eq.~\eqref{eq:spectrum} -- hold throughout the regime, which is thus classically consistent, while a standard particle interpretation and a naive relativistic completion are obstructed there. The thermodynamic route cleanly separates the two statements. Classically the complex poles are poles of the static Green function only: the temporal spectrum \eqref{eq:spectrum} is real and negative in both regimes, and no unitarity question arises. Nor does it arise in the wave-function representation of the fluid~\cite{Van23a,SzucsVan25m}, which complexifies the fluid variables while $\varphi$ remains classical. It arises for a covariant theory with quantized $\varphi$, where six-derivative gravity with complex-conjugate poles has been argued to be unitary in the Lee--Wick sense~\cite{ModestoShapiro2016}; whether the relaxational theory admits such a completion is open.

\emph{Discussion.}  The Second Law, evaluated on a weakly nonlocal state space, does three things.
It selects first-order relaxational dynamics, therefore Ostrogradsky's theorem does not apply, with stability tied to entropy concavity. It delivers the pressure tensor and energy current alongside the field equation, with Newton's force law emerging from the identity \eqref{eq:holo}. It admits the reactive
cross-coupling of Eq.~\eqref{eq:fieldeq}, through which the field is driven by the bulk viscous pressure of the medium -- a term whose astrophysical consequences (a missing-mass
effect generated by the field's own gradient energy, localized where the dissipative gas is) are partially developed in Ref.~\cite{PszoVan24a}. The same construction with other choices of
$\varepsilon$ reproduces self-consistent scalar gravity~\cite{Einstein1912,Giu97a,Franklin2015SelfconsistentSS,Braganca2018} and the AQUAL field equation~\cite{Milgrom83,BekMil84}, so the framework embeds these theories with their hidden thermodynamic compatibility made explicit. This is not a general resolution of the Ostrogradsky problem but indicates an unexplored escape route through Hamiltionian mechanics (\textit{e.g.}~\cite{Morrison1986,GrmelaOttinger1997}).

There are two limitations. The evolution equation does not derive
from an action, so the standard field-theoretic toolbox -- quantization,
particle content, relativistic completion -- does not apply to it in its
present form; and the relaxation pair $(\ell,\tau)$ is constrained only
through the ratio $\ell^2/\tau$. Both point to the same open problem, the
relativistic extension of the thermodynamic construction.

\begin{acknowledgments}
	\emph{Acknowledgements.} We thank R.~Trasarti-Battistoni for insightful remarks. The authors
	acknowledge support by the grant NKFIH NKKP-Advanced 150038 and an STSM
	Grant from FuSe COST Action [CA24101], funded by COST.
    The authors used Claude and ChatGPT to assist with literature synthesis and revising scientific claims. The authors provided the draft and thermodynamic background, carried out all calculations and the physical interpretation and independently reviewed and verified all resulting material.
\end{acknowledgments}

\bibliographystyle{apsrev4-2}
\bibliography{references}

\end{document}